\documentclass[final,3p,times,twocolumn]{elsarticle}

\usepackage{amssymb}
\usepackage{xcolor}
\usepackage{amsthm}

\journal{Physics Letters B}

\begin{document}

\begin{frontmatter}
  


  \title{Experimental investigation of ground-state properties of $^7$H\\ with transfer reactions}

  
  \author[usc]{M.~Caama\~no\corref{cor1}}
  \ead{manuel.fresco@usc.es}
  \address[usc]{IGFAE – Universidade de Santiago de Compostela, E–15706 Santiago de Compostela, Spain}
  \cortext[cor1]{Corresponding author}
  
  \author[ganil]{T.~Roger}
  \ead{roger@ganil.fr}
  \address[ganil]{GANIL, CEA/DSM–CNRS/IN2P3, BP 55027, F–14076 Caen Cedex 5, France}
  
  \author[usevilla]{A.~M.~Moro}
  \address[usevilla]{Universidad de Sevilla, E–41080 Sevilla, Spain}

  \author[ganil]{G.~F.~Grinyer\fnref{alt1}}
  \fntext[alt1]{Present address: Department of Physics, University of Regina, Regina, SK S4S 0A2, Canada}

  \author[ganil]{J. Pancin}

  \author[kvi]{S.~Bagchi\fnref{alt2}}
  \fntext[alt2]{Present address: Indian Institute of Technology (Indian School of Mines), Dhanbad, Jharkhand - 826004, India}
  \address[kvi]{KVI–CART, University of Groningen, NL–9747 AA, Groningen, The Netherlands}

  \author[leuven]{S.~Sambi}
  \address[leuven]{Instituut voor Kernen Stralingsfysica, KU Leuven, B–3001 Leuven, Belgium}

  \author[lpc]{J.~Gibelin}
  \address[lpc]{LPC Caen, Universit\'e de Caen Basse–Normandie–ENSICAEN–CNRS/IN2P3, F–14050 Caen Cedex, France}

  \author[usc]{B.~Fern\'andez–Dom\'inguez}
  
  \author[kyoto]{N.~Itagaki}
  \address[kyoto]{Yukawa Institute for Theoretical Physics, Kyoto University, Kitashirakawa Oiwake–Cho, Kyoto 606–8502, Japan}

  \author[usc]{J.~Benlliure}

  \author[usc]{D.~Cortina–Gil}

  \author[ganil]{F. Farget\fnref{alt3}}
  \fntext[alt3]{Present address: LPC Caen, Universit\'e de Caen Basse–Normandie–ENSICAEN–CNRS/IN2P3, F–14050 Caen Cedex, France}

  \author[ganil]{B.~Jacquot}

  \author[ganil]{D.~P\'erez–Loureiro\fnref{alt4}}
  \fntext[alt4]{Present address: Canadian Nuclear Laboratories, Canada}

  \author[usc]{B.~Pietras}

  \author[leuven]{R.~Raabe}

  \author[usc]{D.~Ramos\fnref{alt5}}
  \fntext[alt5]{Present address: GANIL, CEA/DSM–CNRS/IN2P3, BP 55027, F–14076 Caen Cedex 5, France}

  \author[ganil]{C.~Rodr\'iguez~Tajes}

  \author[ganil]{H.~Savajols}

  \author[ipn]{M.~Vandebrouck\fnref{alt6}}
  \fntext[alt6]{Present address: Irfu, CEA, Universit\'e Paris–Saclay, 91191 Gif–sur–Yvette, France}
  \address[ipn]{IPN Orsay, Universit\'e Paris Sud, IN2P3 – CNRS, F–91406 Orsay Cedex, France}


\begin{abstract}
The properties of nuclei with extreme neutron–to–proton ratios, far from those naturally occurring on Earth, are key to understand nuclear forces and how nucleons hold together to form nuclei. $^7$H, with six neutrons and a single proton, is the nuclear system with the most unbalanced neutron–to–proton ratio known so far. However, its sheer existence and properties are still a challenge for experimental efforts and theoretical models. Here we report experimental evidences on the formation of $^7$H as a resonance, detected with independent observables, and the first measurement of the structure of its ground state. The resonance is found at $\sim$0.7~MeV above the $^3$H+4n mass, with a narrow width of $\sim$0.2~MeV and a $1/2^+$ spin and parity. These data are consistent with a $^7$H as a $^3$H core surrounded by an extended four-neutron halo, with a unique four-neutron decay and a relatively long half-life thanks to neutron pairing; a prime example of new phenomena occurring in what would be the most pure-neutron nuclear matter we can access in the laboratory.
\end{abstract}


\end{frontmatter}

\section{\label{intro} Introduction}

Experimentally, the use of nuclear reactions with unstable nuclei permits to explore regions of the nuclear chart far from stability and beyond the limits of particle binding by systematically adding or removing nucleons~\cite{han87}. This is particularly feasible for light nuclei, where the binding limits and large neutron–to–proton ratios can be reached by adding few nucleons to stable isotopes. In hydrogen, experiments have already obtained evidences of $^{4,5,6,7}$H isotopes, a chain of four resonances beyond the neutron dripline~\cite{coh64,min69,mey79,sen82,fra85,mil86,bel86,bla91,sid03,mei03b,sid04,gur05,tom65,you68,kor01,mei03, gol03,gol04b,ste04,gol05,ter05,wuo17,kor03,caa07,bez20,muz21}. These systems are found to decay by neutron emission and thus believed to be built on a core of $^3$H surrounded by neutrons confined by a centrifugal barrier and grouped in pairs. The systematic study of such a chain helps to understand the evolution of nuclear phenomena, as the effect of neutron pairing and the properties of dilute nuclear matter~\cite{hag07}, away from stability and into the nuclear continuum. At the end of this chain we find the super-heavy $^7$H isotope, the nucleus with the most unbalanced neutron–to–proton ratio of the nuclear chart. $^7$H is expected to display unusual characteristics and open questions. Neutron pairing may render $^7$H the least unstable of the chain despite being the most neutron-rich, forcing it to decay directly to $^3$H by emitting four neutrons simultaneously~\cite{kor03,gol04}. However, other theoretical models predict the resonance to be $\sim$3~MeV above the $^3$H+4n mass~\cite{tim04,aoy09}, and thus favouring a sequential neutron emission over a unique simultaneous decay. Concerning its structure, a recent model based on Antisymmetrised Molecular Dynamics (AMD) describes the four outer neutrons of $^7$H grouped in pairs, and acting as two bosons held together by their interaction with the $^3$H core in a di-neutron condensate~\cite{aoy09}; an exclusive feature of $^7$H among the known hydrogen resonances since a di-neutron condensate needs at least two neutron pairs, hence four valence neutrons. The behaviour of these neutrons can also help to understand the influence and strength of T=3/2 three-body neutron forces, which are key ingredients of the Equation of State that rules the properties and structure of compact stars and supernovae~\cite{akm98,oer17}. The AMD model also predicts that the most probable configuration keeps the di-neutron pairs in a region where the nuclear density of the system drops to 0.02 fm$^{-3}$, almost 10 times lower than the saturation density of nuclear matter. These conditions of density and extreme neutron–to–proton ratio are unique in a nuclear system that can be studied in the laboratory but similar to those expected in a neutron-star crust~\cite{cha08,gul15}.

\begin{figure} [!t]
\begin{center}
\includegraphics[width=\columnwidth]{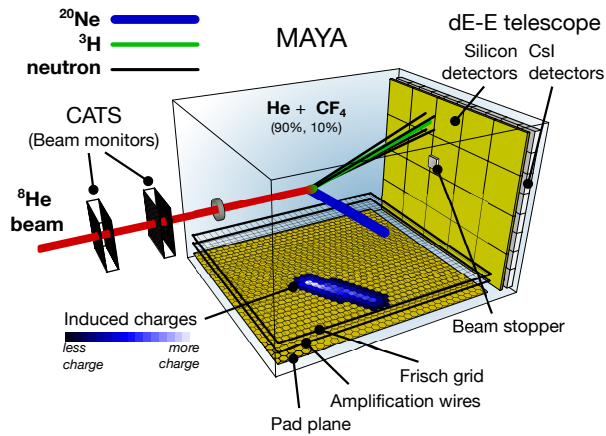}%
\caption{\label{fig1} Schematic drawing of the detection set-up. A typical proton-transfer reaction producing $^7$H with a $^{19}$F nucleus is also shown.}
\end{center}
\end{figure}

From the experimental point of view, the access to these nuclear systems is challenging: their production probabilities are very small; their short lifetimes, of the order of $10^{-22}$~s, prevent direct measurements; and their multi-particle decays complicate the identification of the resonances. This is particularly true for $^7$H, with few experiments with very low statistics reporting evidences on its formation. Meanwhile, its main characteristics, even its existence, are yet to be precisely determined. The first evidence was found in p($^8$He,$^7$H)pp knock–out reactions~\cite{kor03}: a sharp increase in the energy distribution of the beam–like product just above the $^3$H+4n mass was interpreted as hint of the formation of $^7$H. Later, other indications were reported with $^2$H($^8$He,$^7$H)$^3$He reactions: A candidate resonance was found between 1 and 3~MeV above the $^3$H+4n mass~\cite{for07}, while a recent campaign suggests a $^7$H ground state somewhere around 2~MeV and a width below 300~keV and, interestingly, possible excited states at $\sim$6~MeV~\cite{bez20,muz21}. Other experiments using the same reaction channel show less conclusive evidences~\cite{ter07,nik10}. A setup similar to the one presented here was used to identify a resonance peak with a width of $\sim$0.1~MeV, placed at $\sim$0.6~MeV above the $^3$H+4n mass~\cite{caa07} in $^{12}$C($^8$He,$^7$H)$^{13}$N reactions. As to its structure, there are no data on the $^7$H spin and parity.

\begin{figure} [!t]
\begin{center}
\includegraphics[width=\columnwidth]{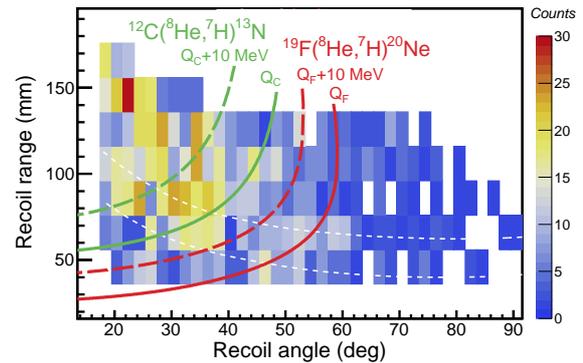}%
\caption{\label{figRangeTheta} Kinematics of target-like recoil. The figure shows the range of target-like products as a function of their recoil angle for events measured in coincidence with a single $^3$H. The colour lines show reference kinematics of one-proton transfer with $^{19}$F (red) and $^{12}$C (green). Q$_{\rm F}$ and Q$_{\rm C}$ correspond to the Q-values of proton-transfer channels forming a $^3$H+4n system with $^{19}$F and $^{12}$C targets, respectively. The white dashed lines enclose the data displayed in Fig.~\ref{fig2}(a).}
\end{center}
\end{figure}

\section{Experimental set-up}
In order to contribute to these sparse results, we have explored binary, one-proton transfer reactions between a $^8$He beam and carbon and fluorine targets to investigate the formation and properties of $^7$H. In 2011, a $^8$He beam with 10$^4$ particles per second was produced and accelerated to $15.4$A~MeV in the SPIRAL facilities at GANIL (France) before being directed to an experimental set-up based on the MAYA active-target~\cite{dem07} and the CATS beam-tracking detectors~\cite{ott99}. The MAYA active-target detector works essentially as a Time, Charge-Projection Chamber where the filling gas also plays the role of reaction target. In this case, we use 176~mbar of a 90\%–10\% molar mix of helium and CF$_4$, with equivalent thickness of 4.2$\cdot$10$^{19}$~atoms/cm$^2$ of $^{19}$F and 1.1$\cdot$10$^{19}$~atoms/cm$^2$ of $^{12}$C. This mixture allowed enough electron multiplication while maintaining a small stopping power of the reaction products. Figure~\ref{fig1} shows a schematic drawing of the experimental setup with a typical $^{19}$F($^8$He,$^7$H)$^{20}$Ne transfer event. 

In a proton-transfer event producing $^7$H, the trajectory of the $^8$He projectile is measured by the CATS beam-tracking detectors before entering MAYA. Once inside, the $^8$He projectile interacts with either a $^{12}$C or $^{19}$F nucleus and transfers a proton, yielding a $^{13}$N or $^{20}$Ne target-like recoil and a $^7$H resonance that, less than $10^{-20}$~s after the reaction, decays into a $^3$H nucleus and four neutrons. The trajectory of the $^{13}$N or $^{20}$Ne target-like is imaged in the segmented pad plane of MAYA, where the angle and range are measured with typical uncertainties of $1.2^{\circ}$ and 16~mm, respectively. The $^3$H scattered at forward angles is identified in a dE–E telescope composed of a first layer of 20 5$\times$5-cm$^2$, 75-$\mu$m thick silicon detectors and a second layer of 80 2.5$\times$2.5 cm$^2$, 1-cm thick CsI crystal detectors. In this setup, neutrons are not detected. Unreacted beam projectiles are collected before the dE–E telescope in a 2$\times$2 cm$^2$ aluminium beam-stopper. The experimental acceptance and uncertainties were verified with the elastic measurement, while the target thickness and beam current were confirmed with the elastic cross-section.

\begin{figure} [!t]
\begin{center}
\includegraphics[width=0.5\columnwidth]{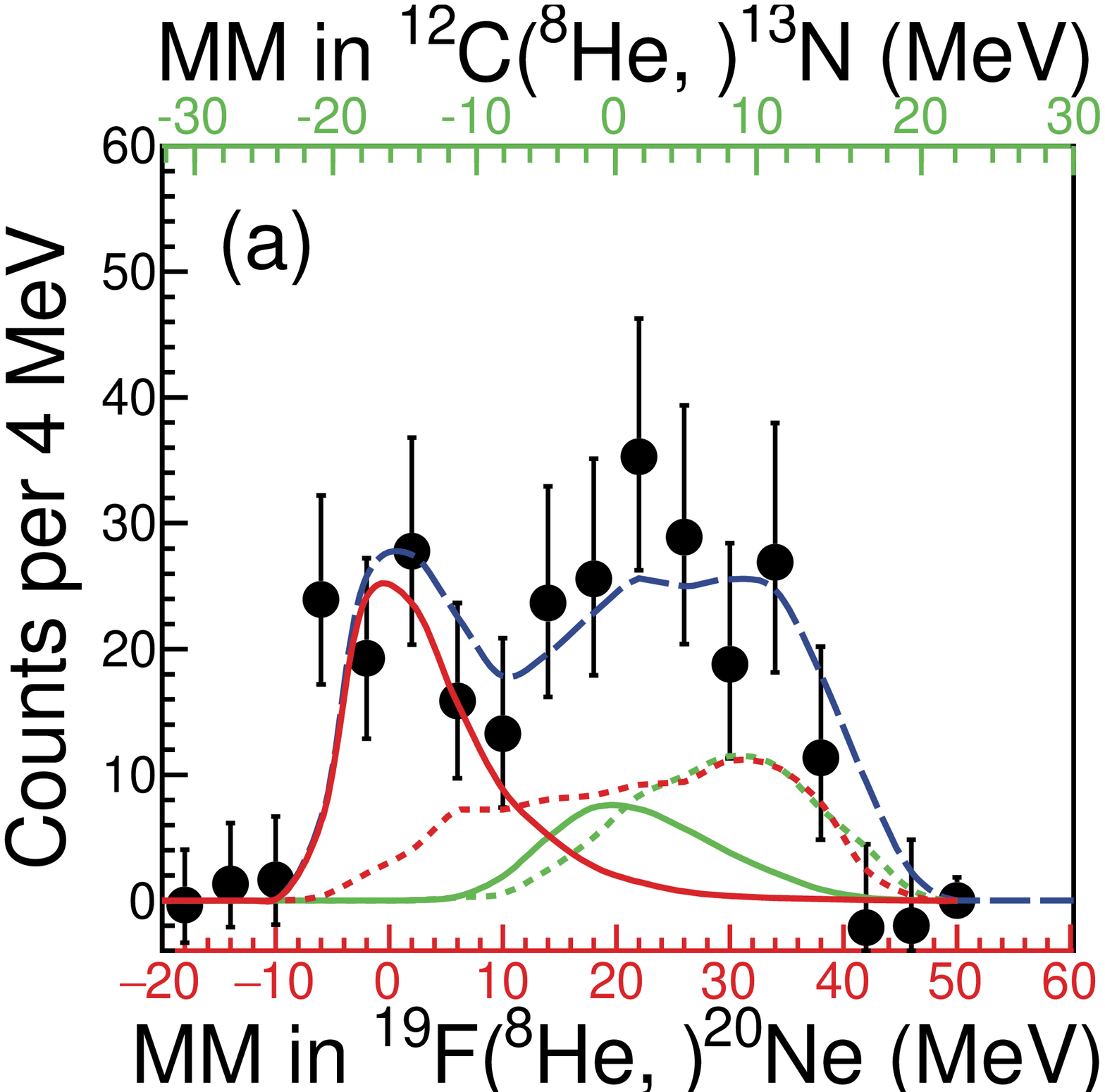}%
\includegraphics[width=0.5\columnwidth]{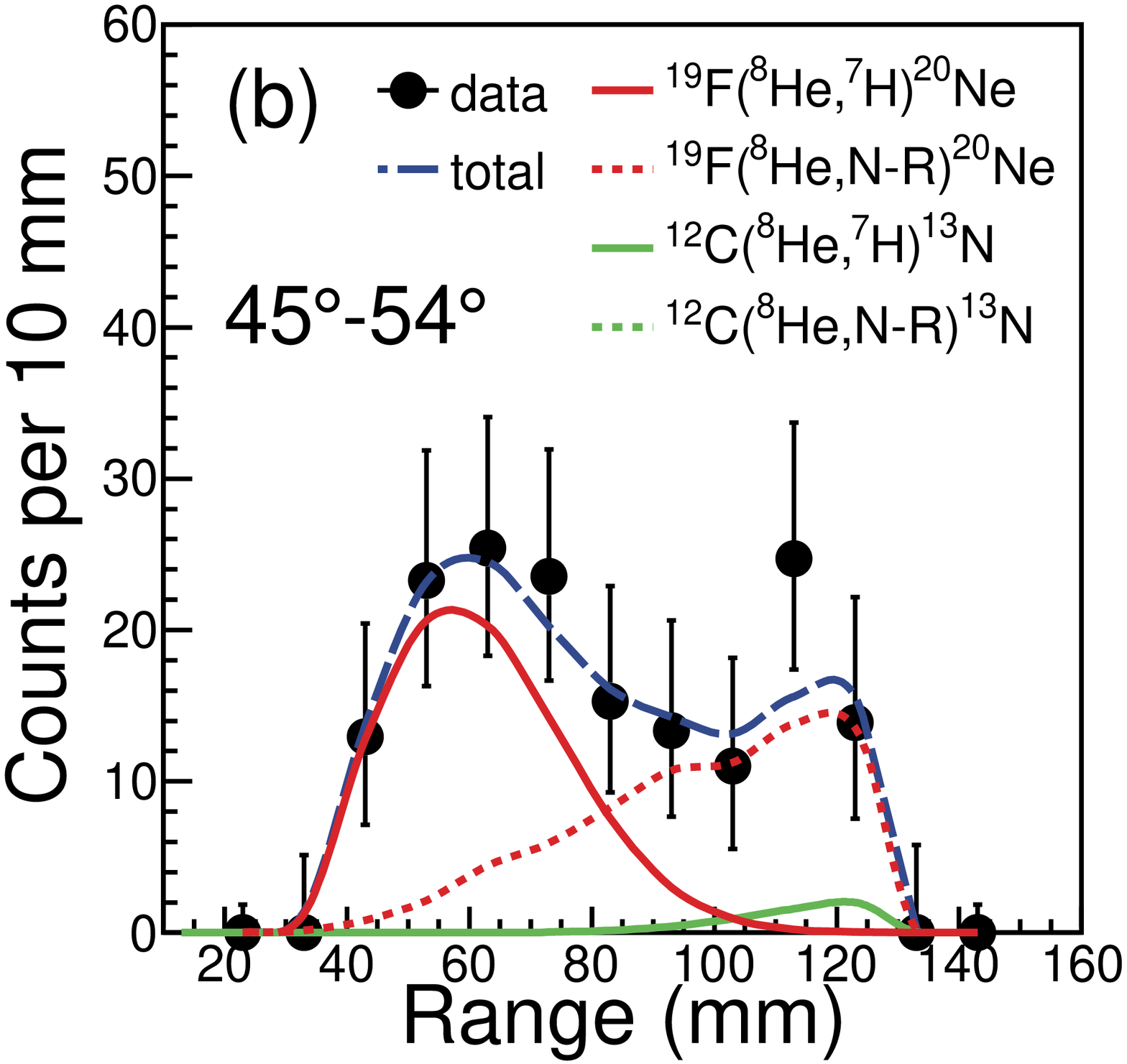}%
\caption{\label{fig2} Experimental evidence of the $^7$H formation. (a) The MM distribution of the data selected between the white dashed lines in Fig.~\ref{figRangeTheta} (black dots) is shown in one-proton transfer reactions with $^{19}$F (red lower axis) or $^{12}$C (green top axis), and after subtraction of background from incomplete events (see text). (b) Measured range distribution of target–like products with a recoil angle between 45$^{\circ}$ and 54$^{\circ}$ in the laboratory frame and after subtraction of background from incomplete events. In both panels, the production of $^7$H (solid lines) and non-resonant backgrounds (\mbox{N-R}, dashed lines) are shown from reactions with $^{19}$F and $^{12}$C targets (red and green lines, respectively).}
\end{center}
\end{figure}

\section{Data Analysis}
\subsection{\label{identification} Identification of $^7$H resonance production}

The selection of $^7$H candidate events is done with the simultaneous measurement of the reaction charged products. A measured proton-transfer reaction with $^{12}$C or $^{19}$F consists of a single $^3$H detected in the dE–E telescope in coincidence with a single trajectory from the target-recoil product, projected on the pad plane. This selection rejects reactions with the helium atoms in the gas, where none of the products ionises enough to induce an image on the pad plane.

Fusion and high-excitation breakup channels with more than two ionising products and/or more than two products on the dE–E telescope are also rejected. However, some breakup events may mimic one-proton transfer if only a $^3$H is detected in the dE–E telescope and a track is recorded in MAYA while the rest of the products are missed. The distribution of these incomplete events is obtained from those detected with a $^3$H in coincidence with any other product in the dE–E telescope and a track inside MAYA. The resulting distribution is subtracted from the set of $^7$H candidate events. This contamination of breakup channels amounts to less than 10\% of single $^3$H and track events, and it follows a smooth behaviour without peaks or recognisable features. Concerning random coincidences between events, they are below 0.1\%.

One-proton transfer reactions correlate the angle and range of the target-like recoils following kinematic lines that depend on the mass of the nuclei involved, including $^7$H. Figure~\ref{figRangeTheta} shows this correlation for the measured data, along kinematic lines corresponding to two reference values for the $^7$H mass. The accumulation of events around the kinematic lines of Q$_{\rm F}$ and Q$_{\rm C}$ in Fig.~\ref{figRangeTheta} is consistent with the production of $^7$H in both channels. A second hint is the abrupt decrease of events we can see around 80 mm along the Q$_{\rm F}$ line. This is due to the angular acceptance in the detection of the scattered $^3$H and it would only affect binary reactions, reinforcing the identification of the accumulation of data around the line as transfer reactions producing $^7$H.

Since $^7$H decays immediately after being formed, its characteristics can be indirectly observed with the missing-mass (MM) method, in which the mass of the undetected participant ($^7$H, in this case) is deduced from the kinematics of the remaining participants in the reaction. Figure~\ref{fig2}(a) shows the MM spectrum of the beam-like product from one-proton transfer reactions with respect to the mass of the $^3$H$+4$n subsystem for data within the largest c.m.~angular region of the $^{19}$F($^8$He,$^3$H+4n)$^{20}$Ne channel that is not cut by acceptance, shown with white dashed lines in Fig.~\ref{figRangeTheta}. Our experimental setup does not perform element separation of the target-like recoils, thus the lower and top horizontal axis of Fig.~\ref{fig2}(a) show the MM as calculated for $^{19}$F and $^{12}$C targets, respectively. The events accumulated around the Q$_{\rm F}$ and Q$_{\rm C}$ kinematic lines shown in Fig.~\ref{figRangeTheta} would appear in Fig.~\ref{fig2}(a) as two peaks: one around zero in the MM spectra of $^{19}$F targets and another around zero for $^{12}$C targets.

The first peak lies in a clean kinematical region: only low-lying states of $^7$H or the lower tail of a 3-body non-resonant (N-R) continuum (which would not be a peak) can populate it. In addition, the angle and energy of the $^3$H detected in coincidence were checked to be within the limits expected from a possible $^7$H decay. Other channels, such as partially reconstructed breakup channels, are subtracted from our selections, as discussed previously. This peak is a safe candidate for the resonant formation of $^7$H with $^{19}$F targets. The peak that would correspond to the formation of $^7$H with $^{12}$C targets is less obvious: it occupies a region also populated by \mbox{N-R} background from both targets and other multi-particle transfer channels. A fit to possible contributions other than $^7$H and its \mbox{N-R} continuum sets an upper limit of 0.2~mb/sr to their production.

\subsection{\label{BW} Resonance mass and half-life}
The values of the resonance mass and width are obtained by fitting a simulation of the main experimental observables to the collected data. The simulation is folded with the experimental resolutions and measurement conditions, and includes the production of $^7$H with both $^{19}$F and $^{12}$C targets, and also N-R events. Besides these channels, other multi-particle transfer reactions were also considered. The measured distributions, and in particular the observed widths, are dominated by the experimental uncertainty, which can reach 10~MeV, depending on the kinematical region, and determines the final uncertainty on the measurement of the resonance parameters.

In order to describe the peak associated with $^7$H, we use a Breit–Wigner probability distribution following the prescription of \cite{caa07}. The mass and width of the resonance, as well as the scaling factor, were treated as free parameters and extracted from a log-likelihood minimisation between the simulation and the measured range distribution in the angular region between $45^{\circ}$ and $54^{\circ}$, shown in  Fig.~\ref{fig2}(b). This region is mainly populated by the production of $^7$H and \mbox{N-R} components, with a very small contribution from the $^{12}$C($^8$He,$^7$H)$^{13}$N channel, making it especially well-suited for a clean fit.

The results of the fit describe a low-lying, narrow resonance state with a mass of $0.73^{+0.58}_{-0.47}$~MeV above the $^3$H+4n mass and a width of $0.18^{+0.47}_{-0.16}$~MeV. The mass value reinforces the notion of $^7$H as the least unstable of the known hydrogen resonances, less than 1~MeV close to being bound, despite being the most neutron-rich. The reinforced stability brought by neutron pairing also gives a narrow width to $^7$H, which translates into a half-life of 5$\cdot 10^{-21}$~s, an order of magnitude longer than the other hydrogen resonances.

Compared to previous experiments, our measured mass and width are in good agreement with the 0.57$^{+0.42}_{-0.21}$ and 0.09$^{+0.94}_{-0.06}$~MeV reported in~\cite{caa07}. The value estimated in~\cite{bez20}, 1.8$\pm$0.5, is larger but not too dissimilar from our result while ref.~\cite{muz21} reports the largest value at 2.2$\pm$0.5~MeV. It is worthy to note that in each of these references, not more than 10 $^7$H events were collected, less than one order of magnitude lower compared to the present work. Evidences of excited states of $^7$H were reported around 6~MeV in refs.~\cite{bez20,muz21}. In the case of the $^{19}$F channel, these states may be hidden in the tail of the peak associated with the $^7$H ground state, although a fit to a tentative state around 6~MeV gives no statistically significant population above 0.1~mb/sr. From the theory side, calculations tend to heavier $^7$H masses. In particular, models based on a di-neutron condensate from AMD~\cite{gol04} and on hyperspherical functions methods~\cite{tim04} describe a resonance with a mass of $\sim$3~MeV~\cite{aoy09}. Concerning the resonance width, ref.~\cite{gol04} finds a strong correlation with its mass and predicts values below 1~keV for masses around 1~MeV, two orders of magnitude below our measured width.\\

\subsection{Angular distribution and structure}
The capabilities of MAYA to measure low energy products and the relatively high intensity of the $^{8}$He beam have allowed us to collect more than 200~events assigned to $^7$H formation, significantly more than previous experiments. The average production cross-section with $^{19}$F is 2.7$\pm 0.5$~mb/sr between 4$^{\circ}$ and 18$^{\circ}$ in the centre of mass reference frame (c.m.), whereas $^{12}$C yields 1.2$^{+0.5}_{-0.6}$~mb/sr between 6$^{\circ}$ and 27$^{\circ}$. Besides the corresponding statistical uncertainty, systematic uncertainties from the number of incoming projectiles and target thickness are around 0.7\%, while the uncertainty on the position and width of the resonance contribute with $\sim$10\% of the cross-section value. The $^{12}$C channel was also measured in \cite{caa07}, reporting 0.04$^{+0.06}_{-0.03}$~mb/sr between 10$^{\circ}$ and 48$^{\circ}$. When evaluated in the same angular region, our measurement averages to 0.4$^{+0.2}_{-0.3}$~mb/sr, a larger value although still compatible within 1.1 of standard deviation. 

Previous experiments mostly used $^2$H($^8$He,$^7$H)$^3$He reactions, obtaining results that vary with the beam energy and angular coverage. At 15.3~AMeV, values below 0.1~mb/sr were measured in a wide 0$^{\circ}$-50$^{\circ}$ region in c.m.~\cite{for07}. Around 25~AMeV, ref.~\cite{ter07} estimates a cross-section below 0.02~mb/sr in 9$^{\circ}$-21$^{\circ}$ in c.m., while ref.~\cite{bez20} reports 0.025~mb/sr between 17$^{\circ}$ and 27$^{\circ}$, and in ref.~\cite{muz21} even seems to increase beyond 0.04~mb/sr for angles below 10$^{\circ}$ in c.m. When increasing the beam energy to 42~AMeV, ref.~\cite{nik10} finds a similar 0.03~mb/sr between 6$^{\circ}$ and 14$^{\circ}$ in c.m. In addition to these transfer measurements, the pioneering work of ref.~\cite{kor03} calculates the production of $^7$H around 0.01~mb/sr$\cdot$MeV with proton knock-out reactions. While a precise comparison is difficult due to the different reaction mechanisms and angular coverage, and their low statistics, the complete list of results seems to suggest a dependence of the cross-section with the target size. An explanation for this behaviour is beyond the scope of this work. However, these results, together with the proposition in ref.~\cite{muz21} of an extreme peripheral character of the $^7$H ground-state population, suggest that a deeper understanding of the reaction mechanism leading to the formation of $^7$H may be needed to reproduce the ensemble of experimental data.

\begin{figure} [!t]
\begin{center}
\includegraphics[width=0.5\columnwidth]{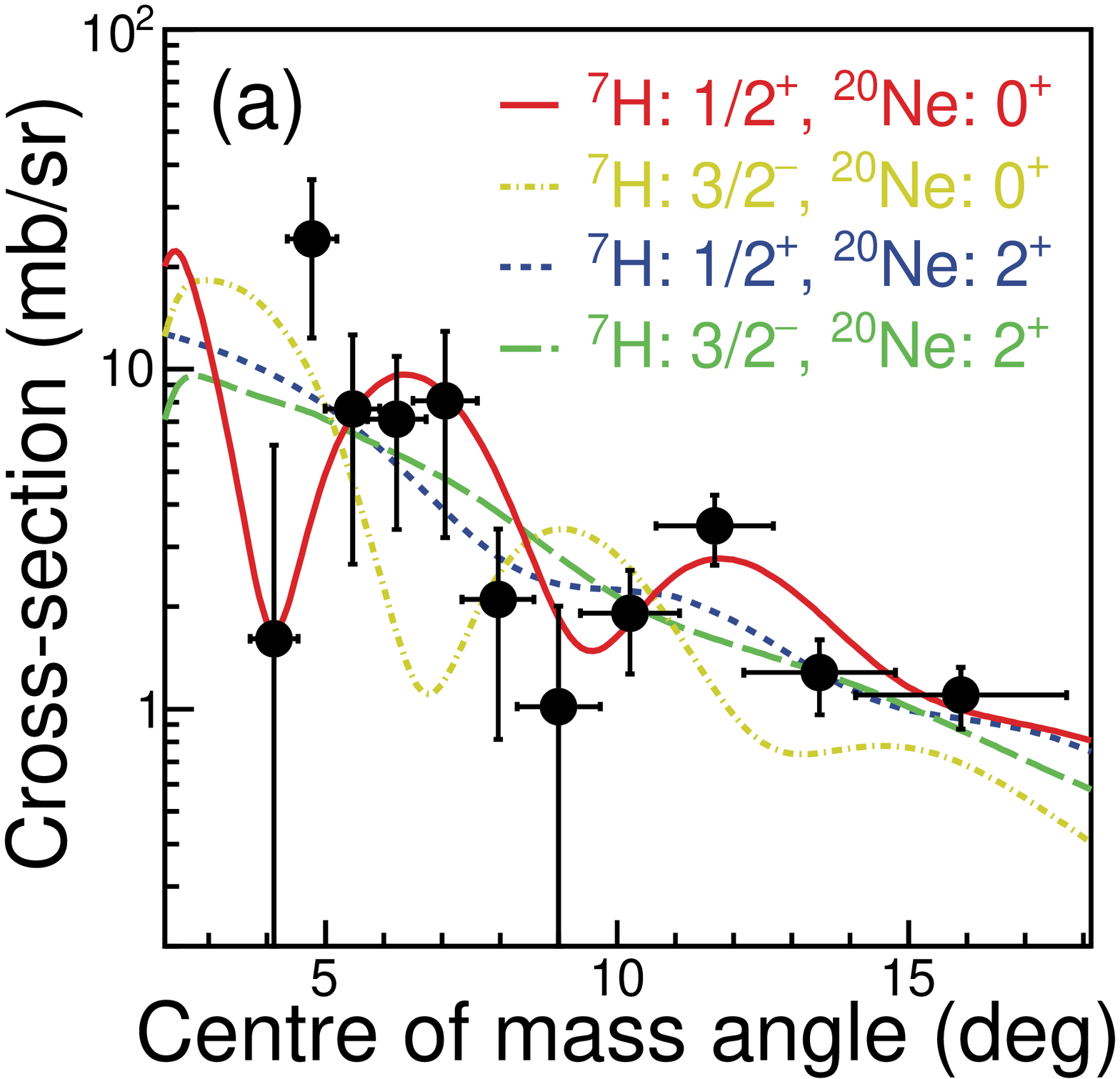}%
 \includegraphics[width=0.5\columnwidth]{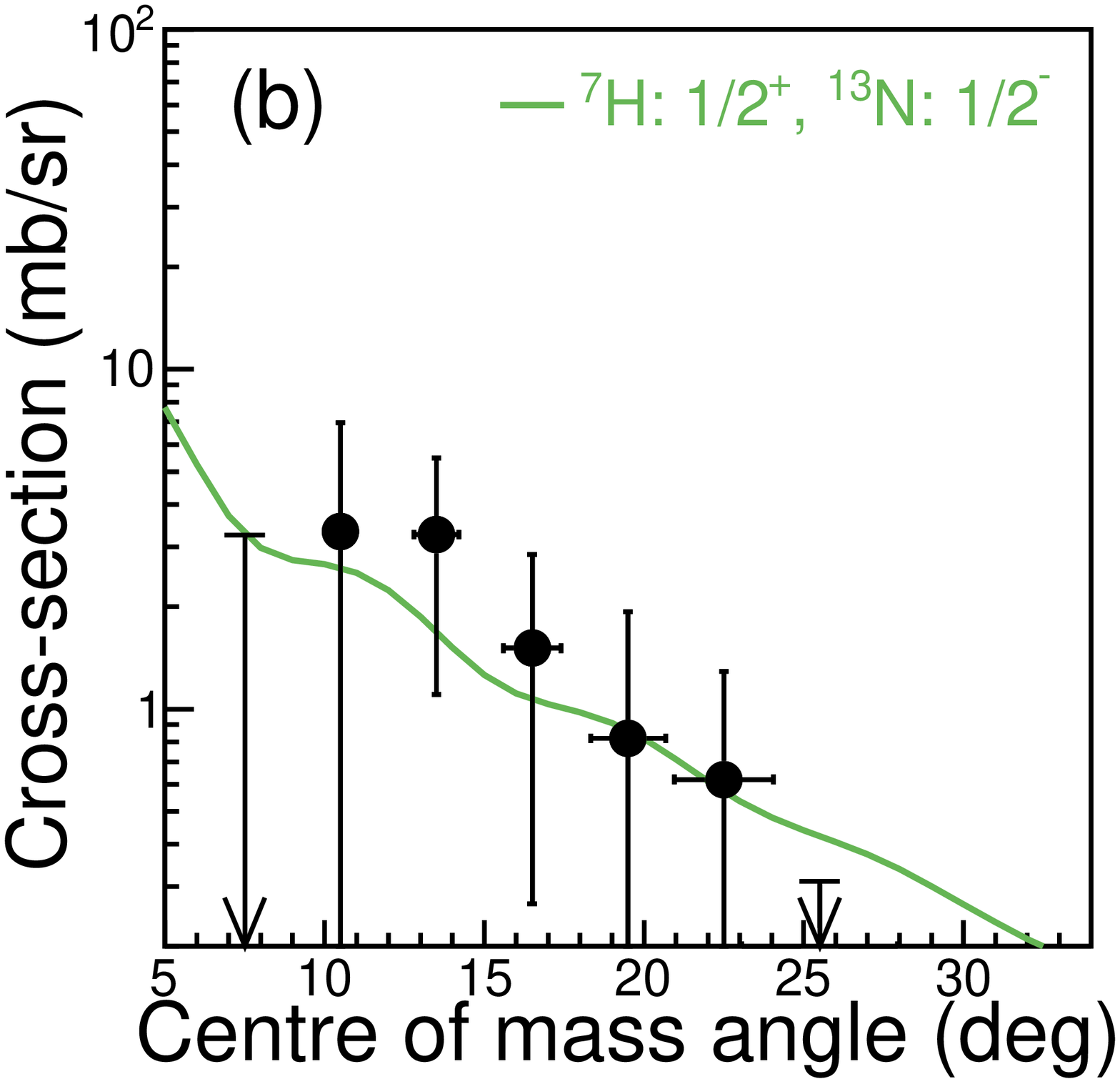}%
\caption{\label{fig4} (a) Measured differential cross-section of the $^{19}$F$+^8$He proton-transfer (black dots) compared with DWBA calculation of a proton transfer to a $1/2^+$ state in $^7$H and a $^{20}$Ne in its $0^+$ ground state (red line) and other spin combinations (yellow, green, and blue lines). (b) Measured differential cross-section of the $^{12}$C$+^8$He proton-transfer channel (black dots) compared with a scaled DWBA calculation performed with a $1/2^+$ $^7$H and the $1/2^{-}$ ground state of $^{13}$N (green line). In both panels, vertical error bars include statistical uncertainty and the effect of the uncertainty in the parameters of the resonance.}
\end{center}
\end{figure}

The improved statistics have also permitted to measure the c.m. angular distribution of the $^7$H production with both targets. Figure~\ref{fig4}(a) shows that the angular distribution of the $^{19}$F($^8$He,$^7$H)$^{20}$Ne channel follows a clear oscillating pattern, with distinct minima. This behaviour is a strong indication of the formation of two well-defined systems in the output channel: $^{20}$Ne and the $^7$H resonance; it offers a further, independent evidence of its production. The measurement of the angular distribution of $^{12}$C($^8$He,$^7$H)$^{13}$N, shown in Fig.~\ref{fig4}(b), suffered from large statistical and systematic uncertainties due to the contribution of competing channels, as discussed previously. The relatively featureless behaviour and its uncertainties do not allow a clear assignment of spin and parity but only a rough assessment of the mean differential cross-section.

The angular distribution is compared in Fig.~\ref{fig4} with different DWBA calculations of the cross-section made with the \textsc{fresco}~\cite{tho88} code. In these calculations, we have used shell-model spectroscopic factors for the $\langle^{20}$Ne$|^{19}$F$\rangle$ overlaps, using the WBT effective interaction by Warburton and Brown~\cite{bwt}. The $^7$H nuclear density was obtained from AMD calculations assuming a di-neutron condensate structure around a $^3$H core~\cite{aoy09}. The resulting cross-sections were folded with the experimental uncertainties and further scaled to match the experimental data. The scaling factor was found to vary between $4.5\pm 2.8$ and $12.7\pm 6.1$, depending on the prescription for the nuclear density of $^8$He. We explore the possibility of populating the $0^+$ ground state or the $2^+$ first excited state of $^{20}$Ne, and for $^7$H, we consider either a $1/2^+$ state with a $^3$H core in its ground state and four outer neutrons, or a $3/2^-$ state with an excited $^3$H core. The relative amplitude of the oscillations and the positions of the minima are best reproduced with a proton transfer to the $0^+$ ground state of $^{20}$Ne and a $1/2^+$ $^7$H resonance. The agreement between the calculations and the data suggests that the di-neutron pairs keep a separation similar to the one predicted by the AMD calculations. Concerning channel mixing, a tentative fit to evaluate a possible mixture of transfer to $0^+$ and $2^+$ states in $^{20}$Ne gives a probability of less than 10\% towards the excited $2^+$ state. The preference for the transfer to the $0^+$ state is surprisingly different from DWBA calculations, which predict a more probable transfer to the $2^+$ state in $^{20}$Ne. The data suggest that either this is not the case or that the resonance has a lower probability of being fully formed in the output channel with the $2^+$ state in $^{20}$Ne. As suggested in ref.~\cite{muz21}, the very radially extended and, thus, very “fragile” nature of the $^{7}$H ground state may play a role to understand this observation and the set of cross-section measurements.

\section{Summary}
In conclusion, we have measured evidences of the formation of the $^7$H resonance with a larger statistical significance than previous attempts via two different reaction channels and performed the characterisation of its ground state with two independent observables: the resonance MM distribution and the angular distribution. From these observables, we have obtained a new determination of the mass and width of $^7$H, and, for the first time, an assignment of spin and parity to its ground state. Together, these results depict the super–heavy $^7$H nucleus as an extended pure–neutron shell around a $^3$H core in a $1/2^+$ ground state. However, the same neutron pairing that allows this large neutron configuration also renders the $^7$H nucleus a long–lived and almost–bound resonance, despite being the system with what would be the largest neutron–to–proton ratio in the nuclear chart known today.

\section*{Acknowledgments}
  The authors thank Navin Alahari for the useful discussions and the careful reading of the manuscript, and Miguel Marqu\'es for his help in the data analysis. The authors are deeply thankful to the technical staff at GANIL for their support and help. This work has been supported by the European Community FP7–Capacities --Integrated Infrastructure Initiative-- contract ENSAR n$^{\circ}$ 262010, and by the Spanish Ministerio de Econom\'ia y Competitividad under contracts FPA2009–14604–C02–01 and FPA2012–39404–C02–01. M.C. acknowledges the support by the Spanish Ministerio de Econom\'ia y Competitividad through the Programmes ``Ram\'on y Cajal'' with the grant number RYC–2012–11585 and ``Juan de la Cierva'' with the grant number JCI2009–05477. A.M.M. is partially supported by the project Ref.~P20\_01247, funded by the Consejer\'{\i}a de Econom\'{\i}a, Conocimiento, Empresas y Universidad, Junta de Andaluc\'{\i}a (Spain) and by “ERDF A way of making Europe."
  
\section*{Competing interests}
The authors declare no competing interests.


  \bibliographystyle{elsarticle-num} 
  \bibliography{caamano}

\end{document}